\RequirePackage{fix-cm}
\documentclass[smallcondensed]{svjour3}     
\smartqed  
\usepackage{graphicx}
\usepackage{amsmath}
\usepackage[colorinlistoftodos]{todonotes}
\usepackage{color}
\begin{document}

\title{Helmholtz Decomposition and Boundary Element Method applied to Dynamic Linear Elastic Problems
\thanks{This work is supported in part by the Australian Research Council through a Discovery Early Career Researcher Award DE150100169 to QS and a Discovery Project Grant DP170100376 to DYCC.}
}


\author{Evert Klaseboer         \and
        Qiang Sun   \and \\
        Derek Y. C. Chan 
}


\institute{E. Klaseboer \at
              Institute of High Performance Computing, 1 Fusionopolis Way, Singapore 138632, Singapore\\
              \email{evert@ihpc.a-star.edu.sg}           
           \and
           Q. Sun \at
              Particulate Fluids Processing Center, Department of Chemical Engineering, The University of Melbourne, Parkville 3010, VIC, Australia \\
              \email{Qiang.Sun@unimelb.edu.au}
           \and 
           D. Y. C. Chan \at
           School of Mathematics and Statistics, The University of Melbourne, Parkville 3010, VIC, Australia and Department of Mathematics, Swinburne University of Technology, Hawthorn 3122, VIC, Australia \\
           \email{D.Chan@unimelb.edu.au}
}

\date{Received: date / Accepted: date}

\titlerunning{Helmholtz decomposition and BEM applied to dynamic linear elasticity}

\maketitle

\begin{abstract}
The displacement field for three dimensional dynamic elasticity problems in the frequency domain can be decomposed into a sum of a longitudinal and a transversal part known as a Helmholtz decomposition. The Cartesian components of both the longitudinal and transverse fields satisfy scalar Helmholtz equations that can be solved using a desingularized boundary element method (BEM) framework. The curl free longitudinal and divergence free transversal conditions can also be cast as additional scalar Helmholtz equations. When compared to other BEM implementations, the current framework leads to smaller matrix dimensions and a simpler conceptual approach. The numerical implementation of this approach is benchmarked against the 3D elastic wave field generated by a rigid vibrating sphere embedded in an infinite linear elastic medium for which the analytical solution has been derived. Examples of focussed 3D elastic waves generated by a vibrating bowl-shaped rigid object with convex and concave surfaces are also considered. In the static zero frequency limit, the Helmholtz decomposition becomes non-unique, and both the longitudinal and transverse components contain divergent terms that are proportional to the inverse square of the frequency. However, these divergences are equal and opposite so that their sum, that is the displacement field that reflects the physics of the problem, remains finite in the zero frequency limit. 


\end{abstract}

\section{Introduction}
\label{intro}

Numerical modeling using dynamic linear elasticity theory has found applications in many fields. It has been used in areas such as geological surveys, earth-soil interaction, sound reduction, crack detection \cite{Iturranan2008} or even in earthquake propagation studies \cite{Beskos1}. Currently, there is renewed interest in this area due to advances in the development of ultrasonic and microfluidic based devices for trapping of biological cells and micro particles \cite{Dual2012}. 

An extensive review of early analytic treatments of the theory of dynamic elasticity is given by Sternberg \cite{Sternberg1960} who, according to Gurtin \cite{Gurtin1973} in his classic survey, tried to introduce the concept of elasticity in ``a form palatable to both engineers and mathematicians''. However, such analytic methods are only suitable for problems with simple geometries, whereas with more general and complex geometries, numerical solutions must be employed.

One of the existing numerical approaches is the finite element method. Although the approach is general, actual implementation can become complicated when domain geometry with regions of different elastic properties are considered, e.g. composite systems with inclusions of different materials. If the geometric properties of the problem necessitate the use of multi-scale grids, spurious refraction or dispersion in wave propagation can arise at the boundaries separating grids of different length scales. In cases where an infinite domain is involved, one further needs to construct the effective outer boundary condition in order to satisfy the Sommerfeld radiation condition at infinity.

Another approach is the boundary element method (BEM) that involves the solution of surface integral equations \cite{CruseRizzo1968,Cruse1968}. Although the resulting matrix system is dense, one only needs to deal with a surface mesh coinciding with the geometry of the domain boundaries thereby reducing a 3D problem to a 2D problem, see for example Rizzo \emph{et al.} \cite{Rizzo1985}, or Beskos \cite{Beskos1,Beskos2}. This approach involves handling of at least weakly singular but integrable kernels in the integral equations \cite{Bu2014}, unless a recently developed desingularization method is employed \cite{SunRoySoc2015}. 

The objective of this paper is to apply the Helmholtz decomposition to dynamic elasticity problems in the frequency domain using the desingiularized boundary element method that provides high precision with fewer number of unknowns or degrees of freedom. The key idea is to use the Helmholtz decomposition of the dynamic elastic equation as described in Landau and Lifshitz \cite{LandauLifshitz} and work directly with the displacement vector field, $\boldsymbol{u}$ which is decoupled into the sum of a transversal field, $\boldsymbol u_T$ and a longitudinal field, $\boldsymbol u_L$. The solution can then be framed in terms of a set of scalar Helmholtz equations that are coupled by given boundary conditions. The divergence free condition on the transversal component and the curl free condition on the longitudinal component can both be framed as Helmholtz scalar equations. Furthermore, these Helmholtz equations, all of the form
\begin{align} \label{eq:Helmholtz} 
 \nabla^2 f + k^{2} f = 0,
\end{align}
with $f$ a scalar function and $k$ the constant wavenumber, can be solved with a recently developed BEM method that does not involve singular integrals \cite{SunRoySoc2015}. 

In conventional BEM applied to Helmholtz equations, it is common practice for the surface to be represented by planar area elements and the unknown functions are taken to be constant within each of these elements. The singularity of the Green's function implies that integrals in which the integration point and the observation point lie in the same area element need to be treated with care. Although the presence of the diverging integrands is an accepted feature of the BEM, it does raise the philosophical question as why a mathematical formulation of physical problems that are well-behaved on boundaries needs to contain mathematical singularities.

In our non-singular version of the BEM \cite{SunRoySoc2015}, the singularities associated with the Green's function are removed analytically so that the surface integrals do not contain diverging integrands. The unknowns are taken to be values of functions at points or nodes that define quadratic surface elements on the boundary. For numerical evaluation of the surface integrals, the value of the integrand at any point within each area element is obtained by quadratic interpolation from the nodal values and such integrals can be evaluated accurately by quadrature. This approach increases the precision over conventional BEM by about 2 orders of magnitude with the same number of degrees of freedom \cite{SunRoySoc2015,KlaseboerEABEM2009}.

It is sometimes  believed that the singular integrals are necessary to create a diagonal dominant matrix after discretizing the integral equations. In theory this is correct, provided that one can calculate the singular terms accurately enough. In practice, however, this almost always leads to considerable errors. For example, for a simple Laplace problem with linear elements, the terms on the diagonal are equal to the sum of the terms off-diagonal \cite{KlaseboerEABEM2009}. Any small error will destroy the critical diagonal dominance. Our non-singular implementation circumvents this difficulty and as a bonus allows us to use higher order elements combined with quadrature to evaluate all integrals. As an additional advantage, it is no longer necessary to calculate the solid angle that simplifies the implementation. 

The theory concerning dynamic linear elasticity is introduced in Sect. \ref{DLE}. A rigid sphere executing harmonic oscillatory motion with a constant amplitude in an infinite linear elastic material will be chosen as a benchmark example. The analytical solution for this problem is given in Sect. \ref{Spheresolution}. Since to the best of our knowledge, it has not been presented elsewhere in the literature, the derivation of this result is sketched in the Appendix. The concept of the desingularized boundary element method is presented in Sect. \ref{BEM}. Some results for the aforementioned vibrating rigid sphere are presented in Sect. \ref{Results} including plots of the displacement field in the 3D domain. Although a simple example has been used as a proof of concept, nevertheless it illustrates the underlying physics and theoretical intricacies. For example, it is found that in the limit of very low wave numbers, each of the decomposed longitudinal and transversal fields will develop a large term of equal magnitude but of opposite sign so that their sum reduces to the correct static solution. Consequently, the BEM framework should be used with caution in the low frequency limit and a discussion of this issue is given in Sect. \ref{Discussion}. For moderate wave numbers these problems do not occur. We also present results for elastic wave pulses generated by an oscillating rigid bowl-shaped object that has both convex and concave surfaces thay can produce focussed elastic waves. Concluding remarks are given in Sect. \ref{Conclusions}.

\section{Dynamic linear elastic waves} 
\label{DLE}
\subsection {The Navier equation}
\label{intro_DLE}
In the time domain, the classical equation of motion without body forces is  
\begin{align} \label{eq:DLE1} 
  \nabla \cdot \boldsymbol{\Sigma} = \rho \frac{\partial^2 \boldsymbol{U}}{\partial t^2},
\end{align}
where the stress tensor, $\boldsymbol{\Sigma}$ and the displacement field, $\boldsymbol{U}$ are functions of position and time, $t$ and $\rho$ is the material density. Assuming a harmonic time variation with angular frequency, $\omega$ for both the stress tensor, $\boldsymbol{\Sigma}=\boldsymbol{\sigma} \; e^{-i\omega t}$ and displacement vector, $\boldsymbol{U}=\boldsymbol{u} \; e^{-i\omega t}$ one obtains, in the frequency domain:
\begin{align} \label{eq:DLE2} 
  \nabla \cdot \boldsymbol{\sigma} = -\rho \omega^2 \boldsymbol{u}.
\end{align}
The infinitesimal strain tensor $\boldsymbol{\epsilon}$ is given in terms of the gradient of $\boldsymbol{u}$ and its transpose:
\begin{align} \label{eq:DLE3} 
\boldsymbol{\epsilon} = \frac{1}{2}(\nabla \boldsymbol{u} + [\nabla \boldsymbol{u}]^T).
\end{align}
For a linear elastic isotropic and homogeneous material,  $\boldsymbol{\sigma}$ and  $\boldsymbol{\epsilon}$ are related by Hooke's Law
\begin{align} \label{eq:DLE4} 
  \frac{\boldsymbol{\sigma}}{2\mu} = \Big[\frac{c^2_L}{2 c^2_T}-1\Big] \; tr(\boldsymbol{\epsilon}) \; \textbf{I} + \boldsymbol{\epsilon},
\end{align}
with $\textbf{I}$ the identity tensor, the trace operator $tr(\boldsymbol{\epsilon}) \equiv \epsilon_{ii}$ (adopting the convention of summation over repeated indices of Cartesian tensors), the constants $c_L$ and $c_T$ are the longitudinal dilatational and transversal shear wave velocities, respectively, that are defined in terms of the Lam\'e constants $\lambda$ and $\mu$ \cite{LandauLifshitz}:
\begin{subequations} \label{eq:cLandcT}
\begin{align}  
  c^2_L &= (\lambda + 2 \mu)/\rho, \\
  c^2_T &= \mu/\rho. 
\end{align}
\label{eq:DLE5}
\end{subequations}
Introducing Eq. \ref{eq:DLE4} into Eq. \ref{eq:DLE2} we obtain two equivalent forms of the Navier equation
\begin{subequations} \label{eq:LongAndTransNavier}
\begin{align}  
  (c^2_L -c^2_T)\nabla (\nabla \cdot \boldsymbol{u}) + c^2_T \nabla^2 \boldsymbol{u} + \omega^2 \boldsymbol{u} = \boldsymbol{0},  \\
   c^2_L\nabla (\nabla \cdot \boldsymbol{u}) - c^2_T \nabla \times \nabla \times \boldsymbol{u} + \omega^2 \boldsymbol{u} = \boldsymbol{0},
\end{align}
\label{eq:DLE7}
\end{subequations}
where Eq. \ref{eq:DLE7}b follows from the identity: $\nabla \times \nabla \times \boldsymbol{u} = \nabla (\nabla \cdot \boldsymbol{u}) - \nabla^2 \boldsymbol{u}$. This result will be the starting point of our subsequent analysis. It will be shown that Eq. \ref{eq:DLE7}b can be used for the analysis of dynamic linear elasticity by applying a Helmholtz decomposition to the displacement field. It turns out that the resulting equations can all be expressed in terms of scalar Helmholtz equations. 

\subsection{The Helmholtz decomposition applied to dynamic linear elasticity}
\label{Theory}
In this section a Helmholtz decomposition will be applied to the Navier equation (Eq. \ref{eq:DLE7}b). It is well known \cite{LandauLifshitz} that the displacement vector $\boldsymbol{u}$ can be decomposed into a transversal and a longitudinal part as
\begin{align} \label{eq:Theory1} 
  \boldsymbol{u}=\boldsymbol{u}_T+\boldsymbol{u}_L,
\end{align}
in which the transversal $\boldsymbol{u}_T$ and the longitudinal  $\boldsymbol{u}_L$ displacements satisfy
\begin{align} \label{eq:Theory2} 
  \nabla \cdot \boldsymbol{u}_T=0,
\end{align}
\begin{align} \label{eq:Theory3} 
  \nabla \times \boldsymbol{u}_L=\boldsymbol{0}.
\end{align}
We now define two wave numbers, one for the transversal component $k_T=\omega / c_T$ and one for the longitudinal component $k_L=\omega/ c_L$ (noting that from Eq. \ref{eq:cLandcT}, $k^2_T>2 k^2_L$). Substituting Eq. \ref{eq:Theory1} into Eq. \ref{eq:DLE7}b and taking into account the conditions of Eqs. \ref{eq:Theory2} and \ref{eq:Theory3}, it can easily be seen that both $\boldsymbol{u}_T$ and $\boldsymbol{u}_L$ satisfy the vector Helmholtz wave equation \cite{LandauLifshitz}:
\begin{align} \label{eq:Theory4} 
  \nabla^2 \boldsymbol{u}_T + k^2_T \boldsymbol{u}_T = \boldsymbol{0},
\end{align}
\begin{align} \label{eq:Theory5} 
  \nabla^2 \boldsymbol{u}_L + k^2_L \boldsymbol{u}_L = \boldsymbol{0}.
\end{align}
These furnish six scalar Helmholtz equations, for each of the $x$, $y$ and $z$ component of the transversal and longitudinal displacements. However, the divergence and curl free conditions of Eq. \ref{eq:Theory2} and Eq. \ref{eq:Theory3} still need to be satisfied separately. It turns out that we can also cast these conditions as additional Helmholtz scalar equations.

\subsection{Longitudinal waves, $\boldsymbol{u}_L$}

\label{longitudinal}
The zero curl condition, Eq. \ref{eq:Theory3}, of the longitudinal part of the displacement vector (also commonly referred to as compression wave), $\boldsymbol{u}_L$ can be satisfied by introducing a scalar potential $\phi$, where 
\begin{align} \label{eq:potential1}
\boldsymbol{u}_L \equiv \nabla \phi. 
\end{align}
Eq. \ref{eq:Theory5} and Eq. \ref{eq:Theory3} can then be replaced by the scalar Helmholtz equation:
\begin{align} \label{eq:Theory7} 
  \nabla^2 \phi + k^2_L \phi = 0.
\end{align}

\subsection{Transversal waves, $\boldsymbol{u}_T$}
\label{transversal}
The zero divergence condition, Eq. \ref{eq:Theory2}, of the transversal part of the displacement vector (a shear wave), $\boldsymbol{u}_T$ can be satisfied by the following general vector identity
\begin{align} \label{eq:EM3} 
  \nabla^2 (\boldsymbol{x} \cdot \boldsymbol u_T) - \boldsymbol{x} \cdot \nabla^2\boldsymbol u_T = 2 \nabla \cdot \boldsymbol u_T,
\end{align}
with $\boldsymbol{x}$ being the position vector: $\boldsymbol{x} = (x,y,z)$. 
Substituting Eqs. \ref{eq:Theory2} and \ref{eq:Theory4} into Eq. \ref{eq:EM3}  gives

\begin{align} \label{eq:Transversal4} 
  \nabla^2 (\boldsymbol{x} \cdot \boldsymbol u_T) + k_T^2 (\boldsymbol{x} \cdot \boldsymbol u_T) = 0.
\end{align}
This is just another Helmholtz equation for the scalar function $(\boldsymbol x \cdot \boldsymbol u_T)$. 
The origin of $\boldsymbol{x}$ can be chosen arbitrarily as can be seen by taking the dot product of a constant vector, $\boldsymbol b$ with Eq. \ref{eq:Theory4} and subtracting this from Eq. \ref{eq:Transversal4}, the result will be a similar equation as Eq. \ref{eq:Transversal4}, but with the vector $\boldsymbol{x}$ replaced by $(\boldsymbol{x}-\boldsymbol{b})$. Thus the transversal part can be described with four scalar Helmholtz equations: one for each of the 3 components of $\boldsymbol u_T$ and one for $(\boldsymbol x \cdot \boldsymbol u_T)$.

Such an approach has been used successfully in electromagnetic scattering problems \cite{Harrington2001} where the electric field $\boldsymbol E$ is divergence free: $\nabla \cdot \boldsymbol{E} = 0$ and satisfies the vector wave equation: $\nabla^2 \boldsymbol{E} + k^{2} \boldsymbol{E} = \boldsymbol{0}$ (interested readers are referred to \cite{KlaseboerIEEE2017,SunPRB2017}).

\subsection{Solution strategy}

To summarize the above findings, the dynamic linear elastic problem can be expressed in terms of four Helmholtz equations with wavenumber $k_T$; three for the $x$, $y$, $z$ components of $\boldsymbol u_T$ (Eq. \ref{eq:Theory4}) and one for the scalar function $(\boldsymbol x \cdot \boldsymbol u_T) $ in Eq. \ref{eq:Transversal4}; and another Helmholtz equation with wavenumber $k_L$ for the longitudinal potential $\phi$ (Eq. \ref{eq:Theory7}). In the current implementation, the Helmholtz equations are solved with a boundary element method, which relates a function on the surface to its normal derivative (see also Sect. \ref{BEM}). In order to retrieve the longitudinal displacement vector $\boldsymbol{u}_L$, the following formula can be employed
\begin{align} \label{eq:Theory8} 
  \boldsymbol{u}_L=\frac{\partial \phi}{\partial n} \boldsymbol{n}
  +\frac{\partial \phi}{\partial t_1} \boldsymbol{t}_1
  +\frac{\partial \phi}{\partial t_2} \boldsymbol{t}_2,
\end{align}
in which $\partial /\partial n \equiv \boldsymbol{n} \cdot \nabla $ is the normal derivative, $\boldsymbol{n}$ is the unit normal vector, $\partial /\partial t_1 \equiv \boldsymbol t_1\cdot \nabla $ and $\partial /\partial t_2 \equiv \boldsymbol t_2\cdot \nabla $ are the two tangential derivatives along the unit tangential vectors $\boldsymbol t_1$ and $\boldsymbol t_2$ on the surface.  

Essentially, the above described approach is a combination of the soundwave scalar Helmholtz solution for longitudinal waves of Sect. \ref{longitudinal} (see also \cite{SunRoySoc2015}) and the transversal wave approach similar to the one used in electromagnetic scattering (for more details see \cite{KlaseboerIEEE2017} and \cite{SunPRB2017}).

\section{An analytical solution for a vibrating sphere}
\label{Spheresolution}

The analytical solution for a radially oscillating sphere as described in Lautrup \cite{Lautrup} is well known but unfortunately it is less suitable as a numerical test case, since the transversal component is zero due to symmetry considerations. 

Here we consider the waves generated in an elastic medium surrounding a rigid sphere with radius $a$, with the origin of the coordinate system located at the center of the sphere. The sphere executes harmonic displacement of constant amplitude so that in the frequency domain, the prescribed displacement on the surface of the sphere is $\boldsymbol{u}=\boldsymbol{u}_0$, with $\boldsymbol u_0$ a constant vector. The $i^{th}$ component ($i=x,y,z$) of the analytical solution for such a case is (see Appendix for derivation)

\begin{equation} \label{eq:Analytic1} 
\begin{aligned}
u_i ={} &   c_1   \Big[e^{ik_Tr}\big[1+G(k_Tr)\big] - e^{ik_Lr}\frac{k_L^2}{k_T^2} G(k_Lr)\Big] \frac{2a}{r} u_i^0\\
      & +c_1 \Big[e^{ik_Tr} F(k_Tr) - e^{ik_Lr}\frac{k_L^2}{k_T^2}F(k_Lr)\Big] \frac{2a}{r^3} x_i (x_ju_j^0)\\
      & -c_2 e^{ik_Lr} \Big[\delta_{ij} (ik_Lr-1) +x_i x_j k_L^2F(k_Lr)\Big] \frac{a^3}{r^3} u_j^0,
\end{aligned}
\end{equation}
where $r$ is the radial coordinate, $\delta_{ij}$ is the Kronecker delta function and the Einstein summation convention is taken over repeating indices. The functions $F(x)$ and $G(x)$ are defined as
\begin{equation} \label{eq:Analytic2} 
F(x) = -1-\frac{3i}{x}+\frac{3}{x^2},
\end{equation}
\begin{equation} \label{eq:Analytic3} 
G(x) = \frac{i}{x}-\frac{1}{x^2}.
\end{equation}
The terms proportional to $e^{ik_Tr}$ correspond to the divergence free transversal part and the terms proportional to $e^{ik_Lr}$ correspond to the curl free longitudinal part. The constants $c_1$ and $c_2$ can conveniently be expressed in terms of four other constants $A$, $B$, $C$ and $D$ as $c_1 = -B/(DA-BC)$ and $c_2 = A/(DA-BC)$ that are defined as: 
\begin{subequations}
\begin{align} \label{eq:AnalyticA} 
  A &= 2e^{ik_Ta}F(k_Ta) - 2e^{ik_La} (k_L/k_T)^2 F(k_La),  \\
 \label{eq:AnalyticB} 
  B &= -e^{ik_La} (k_L a)^2 F(k_La), \\
 \label{eq:AnalyticC} 
  C &= 2e^{ik_Ta}\Big[1+G(k_Ta)\Big] - 2e^{ik_La} (k_L/k_T)^2 G(k_La), \\
 \label{eq:AnalyticD} 
  D &= -e^{ik_La}\big( ik_La -1 \big). 
\end{align}
\end{subequations}

The method of constructing the solution in Eq. \ref{eq:Analytic1} is outlined in the Appendix. However, it can be verified by direct substitution that Eq. \ref{eq:Analytic1} is indeed a solution of the Navier equation with the boundary condition $u_i=u_i^0$ on the surface $r=a$ and it decays for large values of $r$. 

Perhaps also worth mentioning, although we will not use it in the current work, is the solution that corresponds to the zero tangential stress boundary condition. That is, the boundary condition $u_i=u_i^0$ is replaced by $(\sigma_{ij}n_j)t_i=0$ and $u_in_i=u_i^0n_i$ on the surface of the sphere. The constants $A$, $C$ and $D$ that appear in the coefficients $c_1$ and $c_2$ in Eq. \ref{eq:Analytic1} then have to be replaced by $A'$, $C'$ and $D'$ ($B$ remains the same) as
\begin{subequations}
\begin{align} 
  A' &=  e^{ik_Ta}(ik_Ta - 1) + A, \\
  C' &=  C + A, \\
  D' &=  D + B.
\end{align}
\label{eq:AnalyticACD'} 
\end{subequations}
\section{Desingularized boundary element method for Helmholtz problems}
\label{BEM}
In Sects. \ref{Theory} - \ref{transversal} it was shown that the problem of dynamic linear elasticity can be expressed in terms of five scalar Helmholtz equations in the form of Eq. \ref{eq:Helmholtz}: four of them with wavenumber $k_T$ (Eqs. \ref{eq:Theory4} and \ref{eq:Transversal4}) and another one with wavenumber $k_L$ (Eq. \ref{eq:Theory7}). Here it will be shown how a scalar Helmholtz equation can be solved efficiently using the framework of the boundary element method. The boundary element method has the advantage that only values of the unknown function on  boundaries, $S$, need to be found, and from which values anywhere in the 3D domain can be calculated. In the context of Helmholtz equations, a further advantage of the boundary element method is the fact that the Sommerfeld radiation condition at infinity is automatically satisfied. Thus the boundary element method is especially suited for an object embedded in an infinite domain. Some recent advances in the boundary element method include the concept of full desingularization \cite{KlaseboerJFM2012}, which allows for high accuracy with reduced implementation effort. 

The classical boundary element method is expressed as (see for example Becker \cite{Becker1992}, Kirkup \cite{Kirkup1998} or any classical textbook on boundary element methods)
\begin{align}\label{eq:BEM1} c(\boldsymbol x_0)\phi (\boldsymbol x_0)  + \int_S \phi(\boldsymbol x) \frac{\partial H} {\partial n} \text{ d}S(\boldsymbol x) = \int_S   \frac{\partial \phi(\boldsymbol x)} {\partial n} H \text{ d}S(\boldsymbol x)\end{align}
in which the Green's function for the Helmholtz equation is defined as $H\equiv H(\boldsymbol x, \boldsymbol x_0)=e^{ikr}/r$, with $k$ the wavenumber, $r=|\boldsymbol x - \boldsymbol x_0|$, and $\boldsymbol x_0$ and $\boldsymbol x$ the observation and integration points, respectively. The variable $c(\boldsymbol x_0)$ is the solid angle when $\boldsymbol x_0$ is on the boundary and $c=4\pi$ when $\boldsymbol x_0$ is situated in the domain. The boundary element method relates the potential $\phi$ to its normal derivative $\partial \phi / \partial n$, where $\partial / \partial n \equiv \boldsymbol n \cdot \nabla$ (the unit normal vector on the surface $S$ is $\boldsymbol n = \boldsymbol n(\boldsymbol x)$ and points out of the domain). If, for example, $\phi$ (or $\partial \phi / \partial n$) is specified as a given boundary condition, then Eq. \ref{eq:BEM1} can be solved for $\partial \phi / \partial n$ (or $\phi$). If the surface $S$ is discretized into $N$ nodes, Eq. \ref{eq:BEM1} can be written with respect to each node (corresponding to a different $\boldsymbol x_0$), and after the surface integrals are evaluated then results in a $N \times N$ linear matrix system to be be solved numerically. 

A relatively new concept, first introduced by Klaseboer et al. \cite{KlaseboerEABEM2009} is to replace $\phi(\boldsymbol x)$ in Eq. \ref{eq:BEM1} by a known analytical function $\chi(\boldsymbol x)$ that also satisfies the Helmholtz equation, so that 
\begin{align}\label{eq:BEM2} c(\boldsymbol x_0)\chi (\boldsymbol x_0)  + \int_S \chi(\boldsymbol x) \frac{\partial H} {\partial n} \text{ d}S(\boldsymbol x) = \int_S   \frac{\partial \chi(\boldsymbol x)} {\partial n} H \text{ d}S(\boldsymbol x).\end{align}
In addition, $\chi(\boldsymbol x)$ can be constructed to have the following properties
\begin{eqnarray}
  \lim_{\boldsymbol{x}\to \boldsymbol x_0} \chi(\boldsymbol{x}) &=&\phi (\boldsymbol x_0)\label{eq:Lim1} \\ 
 \lim_{\boldsymbol{x}\to \boldsymbol x_0} \frac{\partial\chi(\boldsymbol{x})}{\partial n} &=& \frac{\partial\phi(\boldsymbol x_0)}{\partial n}\label{eq:Lim2}
\end{eqnarray} 
so that when Eq. \ref{eq:BEM2} is subtracted from Eq. \ref{eq:BEM1}, a fully desingularized boundary element method will emerge \cite{KlaseboerJFM2012,SunRoySoc2015}:
\begin{align}\label{eq:BEM3}  
\int_S \left[\phi(\boldsymbol x) -\chi(\boldsymbol x) \right] \frac{\partial H} {\partial n} \text{ d}S(\boldsymbol x) = \int_S   \left[\frac{\partial \phi(\boldsymbol x)} {\partial n}-\frac{\partial \chi(\boldsymbol x)} {\partial n} \right] H \text{ d}S(\boldsymbol x).
\end{align}
Conveniently, the term with the solid angle $c(\boldsymbol x_0)$ no longer appears in Eq. \ref{eq:BEM3}. In this work, we can take
\begin{eqnarray}\label{eq:FuncHelm1} 
\chi(\boldsymbol{x}) &=& \phi(\boldsymbol x_0) \cos{y} +\frac{1}{k}\frac{\partial \phi (\boldsymbol x_0)} {\partial n} \sin{y}, \\ 
y &=& k \; \boldsymbol{n}(\boldsymbol x_0) \cdot (\boldsymbol{x} -\boldsymbol x_0), 
\end{eqnarray} 
so that Eq. \ref{eq:BEM3} can then be written in full as:
\begin{equation}
\begin{aligned}\label{eq:BEM4}  4\pi \phi (\boldsymbol x_0) + \int_S \Big\{ \phi (\boldsymbol{x}) - \phi(\boldsymbol x_0) \cos{y} +\frac{1}{k}\frac{\partial \phi (\boldsymbol x_0)} {\partial n} \sin{y}\Big\} \frac{\partial H} {\partial n} \text{ d}S(\boldsymbol x) = \qquad  \qquad \\ 
\int_S  \Big\{ \frac{\partial \phi (\boldsymbol{x})} {\partial n}-
\frac{\partial \phi (\boldsymbol x_0)}{\partial n} [\boldsymbol{n}(\boldsymbol x_0)\cdot \boldsymbol{n}(\boldsymbol{x})] \cos{y}  +k [\boldsymbol{n}(\boldsymbol x_0)\cdot \boldsymbol{n}(\boldsymbol{x})]  \phi(\boldsymbol x_0) \sin{y} 
\Big\} H \text{ d}S(\boldsymbol x).\end{aligned}
\end{equation}
Note that the terms with $\cos y$ perform the actual desingularization since $y$ tends towards zero as $\boldsymbol x$ approaches $\boldsymbol x_0$, which cancels out the $1/r$ singularity caused by the Green's function $H$ and its normal derivative. Also $\boldsymbol n(\boldsymbol x_0) \cdot \boldsymbol n(\boldsymbol x)$ tends towards unity when $\boldsymbol x$ approaches $\boldsymbol x_0$. In Eq. \ref{eq:BEM4} the terms with $\phi(\boldsymbol x_0)$ and $\partial \phi(\boldsymbol x_0)/ \partial n$ will end up on the diagonal of a resulting matrix system after a discretisation and numerical Gaussian integration has been performed. The term with $4\pi \phi (\boldsymbol x_0)$ originates from the fact that the choice of Eq. \ref{eq:FuncHelm1} when put into Eq. \ref{eq:BEM2} will cause a contribution from the surface at infinity, which turns out to be exactly $4\pi \phi (\boldsymbol x_0)$. This term is only present for external problems (such as the ones described in the current work) and should be omitted for internal problems. 

This framework is free of any weak, strong or hyper singularities associated with the usual implementation of the boundary element method in dynamic linear elasticity. Simple Gauss quadratures can therefore be employed to evaluate integrals over each element including the previously singular ones. In the current implementation, integration over quadratic six noded triangular elements was used with quadratic shape functions \cite{KlaseboerIEEE2017}.

The normal derivative of $(\boldsymbol{x} \cdot \boldsymbol u_T)$ can be expressed in terms of the normal component of $\boldsymbol u_T$ and the dot product of $\boldsymbol x$ with the normal derivative of $\boldsymbol u_T$ as:
\begin{align} \label{eq:xuTderivative}
\frac{\partial (\boldsymbol{x} \cdot \boldsymbol u_T)}{\partial n} & = \boldsymbol u_T \cdot \boldsymbol n + \boldsymbol x \cdot \frac{\partial \boldsymbol u_T}{\partial n} \nonumber \\
& =  u_{Tx} n_x +  u_{Ty} n_y +u_{Tz} n_z 
+ x \frac{\partial  u_{Tx}}{\partial n}
+ y \frac{\partial  u_{Ty}}{\partial n}
+ z \frac{\partial  u_{Tz}}{\partial n}
\end{align}

\begin{figure*}[t]
  \includegraphics[width=0.5\textwidth]{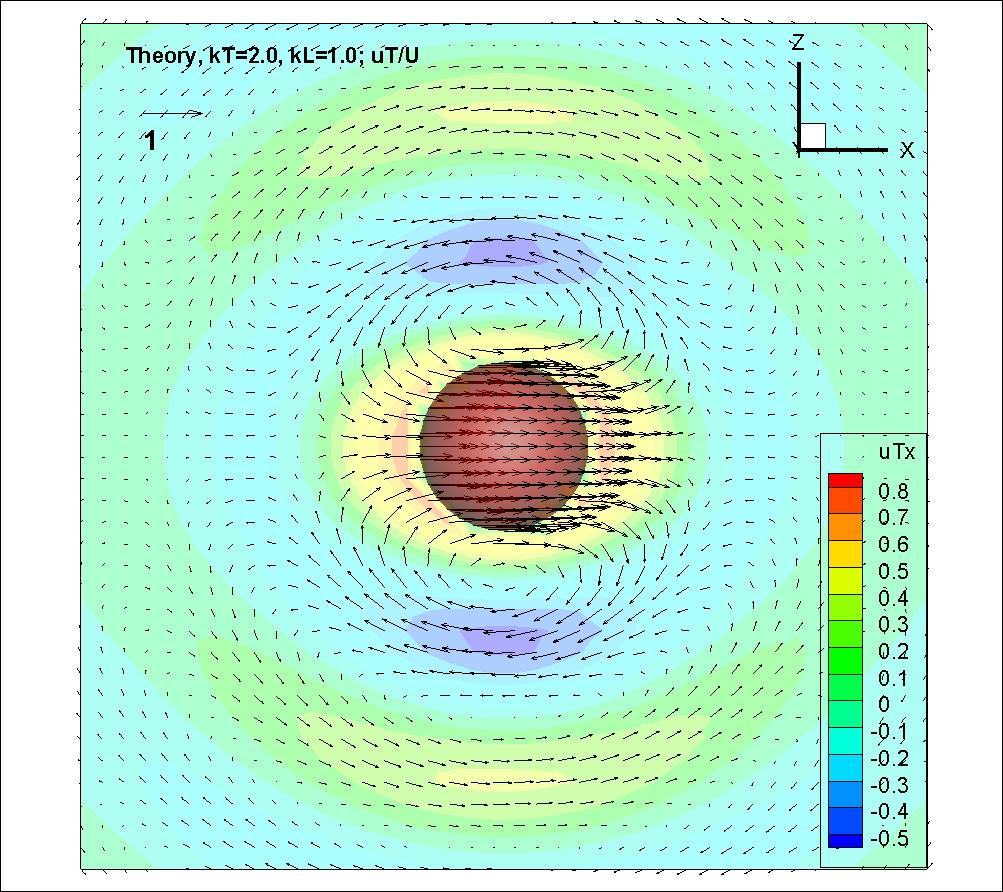}
   \includegraphics[width=0.5\textwidth]{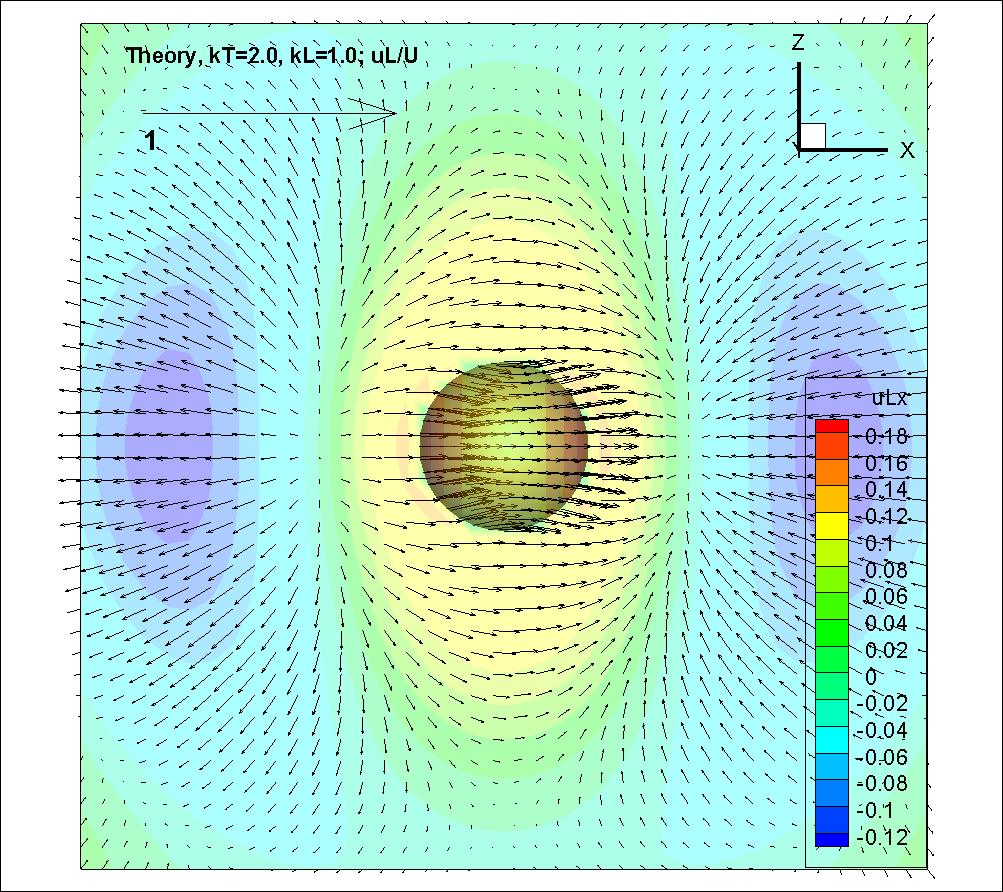}
   \includegraphics[width=0.5\textwidth]{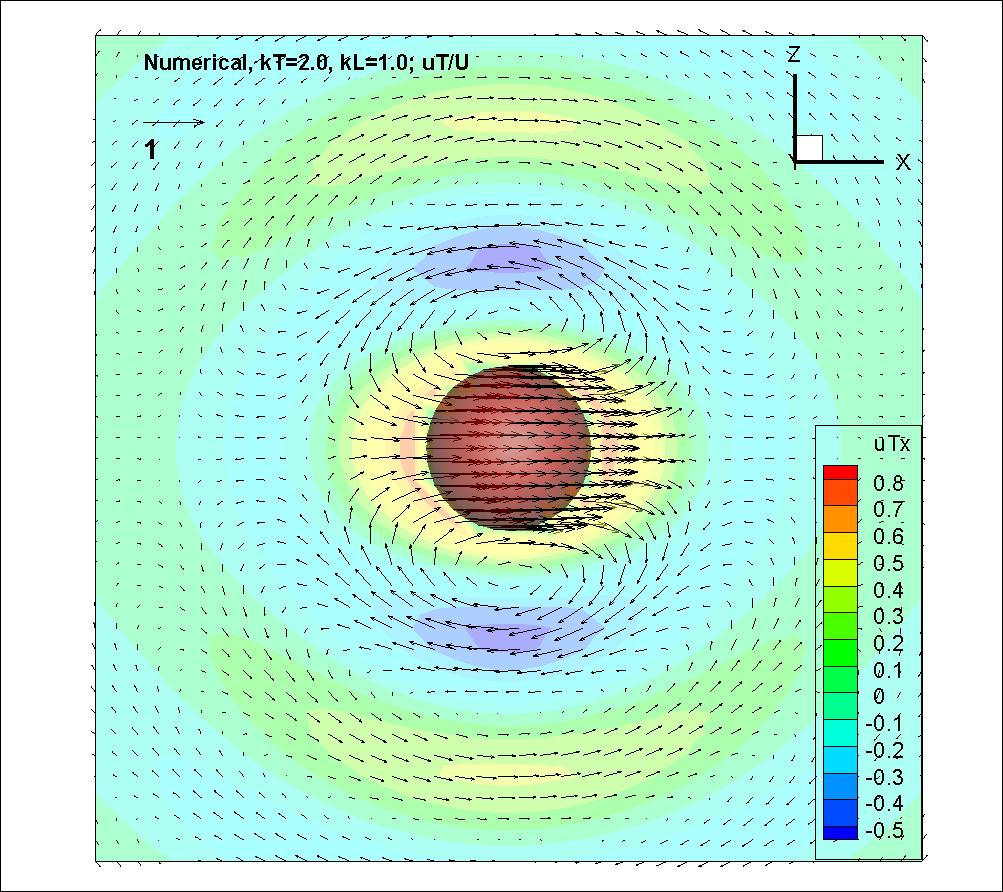}
\includegraphics[width=0.5\textwidth]{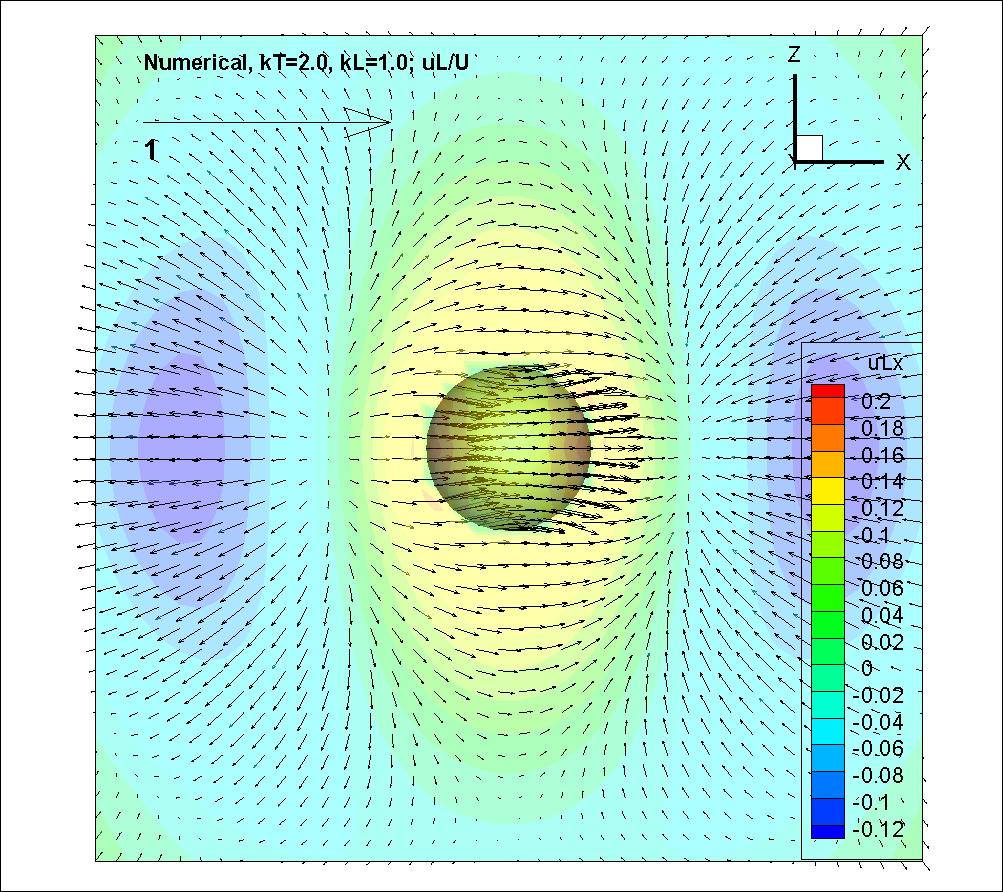}
\caption{Sphere with radius $a$ vibrating in the $x$-direction defined by the boundary condition: $\boldsymbol u^{0} = (U,0,0)$ on the sphere surface $r = a$ at $k_Ta=2.0$ and $k_La=1.0$. Surface and field plot of the displacement field vector $\boldsymbol u$ scaled by $U$. Analytical results for $\boldsymbol u_T$ and $\boldsymbol u_L$ are given in the top left and right images and corresponding numerical BEM results are shown in the lower images. For this particular case $\boldsymbol u_T$ is the dominant term. For corresponding movies to these figures see Sect. \ref{CompMat}.
}
\label{fig:1}       
\end{figure*}

The tangential derivatives in Eq. \ref{eq:Theory8} were calculated using the average of the tangential derivatives on each neighboring element of a node. In the current implementation we used an iterative method with an LU-decomposition framework, such that effectively only two $N \times N$ matrix systems need to be solved (one for $k_T$ and one for $k_L$). To start the iterative process, an estimation for the normal component of the transversal displacement is $u_{Tn}^m = \boldsymbol u_T \cdot \boldsymbol n$ is assumed (for the first iteration, $m=1$ and  $u_{Tn}^1 = 0$). Then, for the next iteration, the normal derivative of the potential is calculated as
\begin{align}\label{eq:Iter1} 
\frac{\partial \phi^{m+1}}{\partial n} = (1-\alpha)\frac{\partial \phi^{m}}{\partial n} + \alpha[\boldsymbol u_0 \cdot \boldsymbol n - u_{Tn}^m],
\end{align}
where a relaxation factor $\alpha$ was used. With the boundary element method (for $k_L$) an estimation for $\phi^{m+1}$ can now be found. Its tangential derivatives in the $\boldsymbol{t}_1$ and $\boldsymbol{t}_2$ direction can be calculated and $\boldsymbol u_L^{m+1}$ is given by Eq. \ref{eq:Theory8}. Since on the boundary $\boldsymbol u_T = \boldsymbol u_0 - \boldsymbol u_L$, with $\boldsymbol u_0 =(U,0,0)$ prescribed, the transversal vector $\boldsymbol u_T^{m+1}$ can be obtained. $\boldsymbol u_T^{m+1}$ is then decomposed into its $x$, $y$ and $z$ components, and, for each component, we apply the boundary element method (now for $k_T$) to get $\partial u_{Tx}^{m+1}/\partial n$, $\partial u_{Ty}^{m+1}/\partial n$ and $\partial u_{Tz}^{m+1}/\partial n$. To satisfy the last Helmholtz equation corresponding to Eq. \ref{eq:Transversal4}, the scalar $\boldsymbol x \cdot \boldsymbol u_T^{m+1}$ is given and its normal derivative is calculated with the boundary element method (again for $k_T$). Since  $\partial u_{Tx}^{m+1}/\partial n$, $\partial u_{Ty}^{m+1}/\partial n$ and $\partial u_{Tz}^{m+1}/\partial n$ are already known, with the help of Eq. \ref{eq:xuTderivative}, a new estimate for $u_{Tn}^{m+1}$ can be obtained. Then the iterative loop can be repeated until convergence is obtained. There are alternative approaches to solve the system of equations, some discussion on such solutions will be presented in Sect. \ref{Results}

\section{Results}
\label{Results}

\begin{figure*}[t]
  \includegraphics[width=0.5\textwidth]{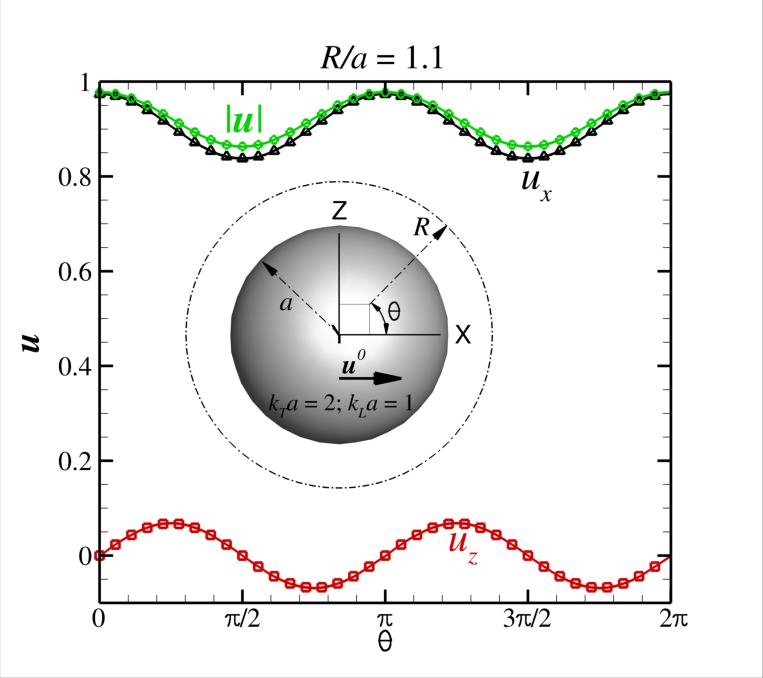}
   \includegraphics[width=0.5\textwidth]{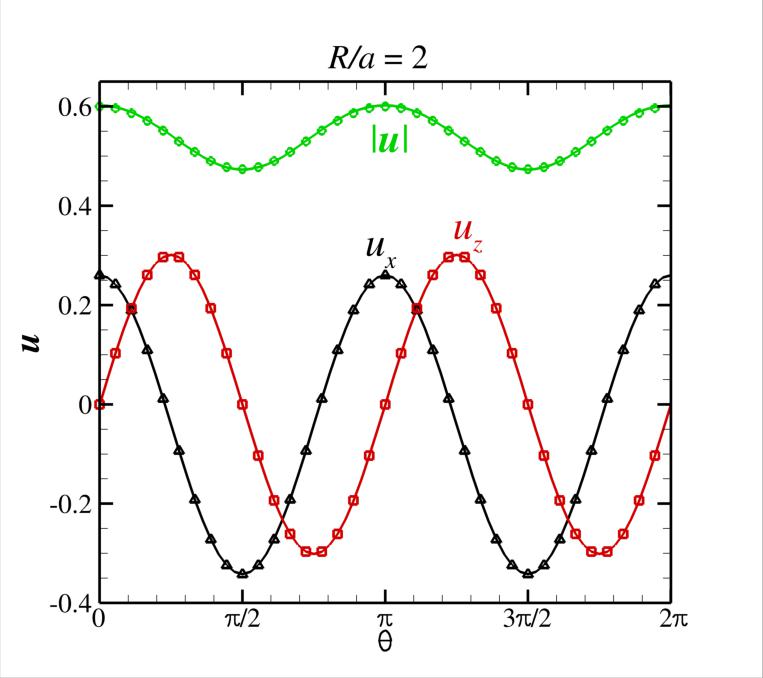}
\caption{Same as in Fig.\ref{fig:1}, field displacement $\boldsymbol u$ scaled by $U$ along circles with radii (left) $R/a=1.1$ and (right) $R/a=2$ on plane $y=0$. Lines: analytic solutions; symbols: numerical solutions.
}
\label{fig:1_1}       
\end{figure*}

\begin{figure*}[t]
  \includegraphics[width=0.5\textwidth]{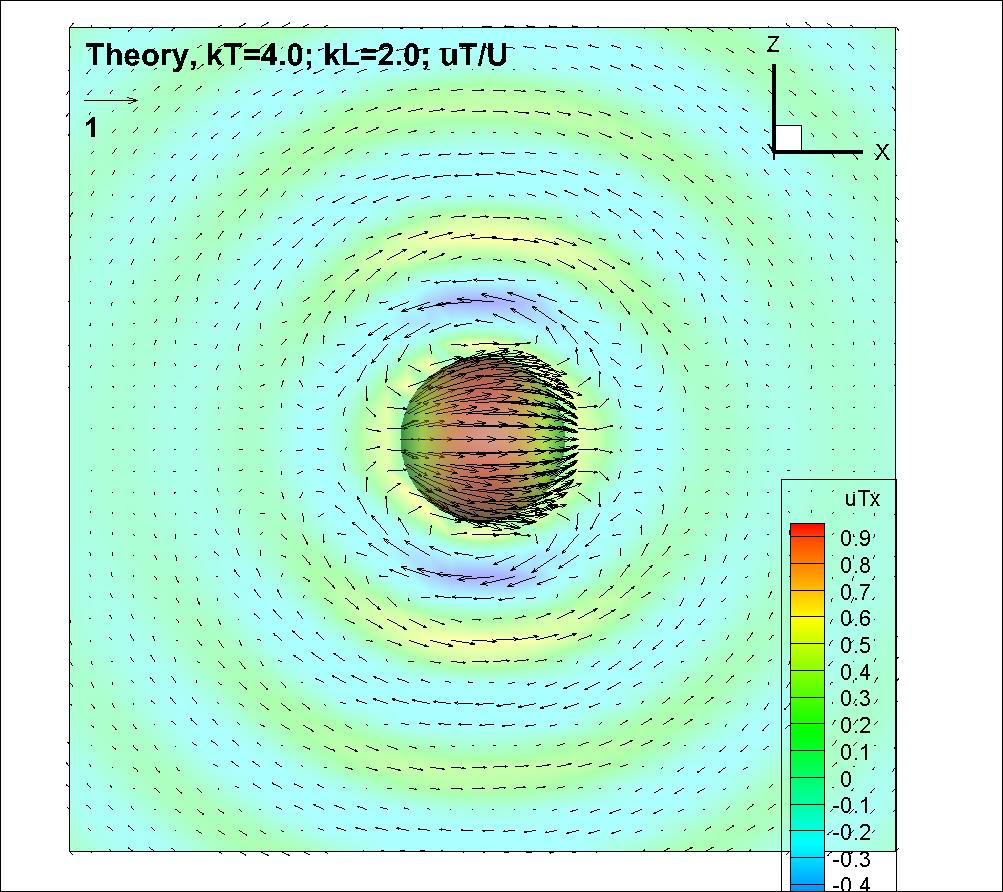}
   \includegraphics[width=0.5\textwidth]{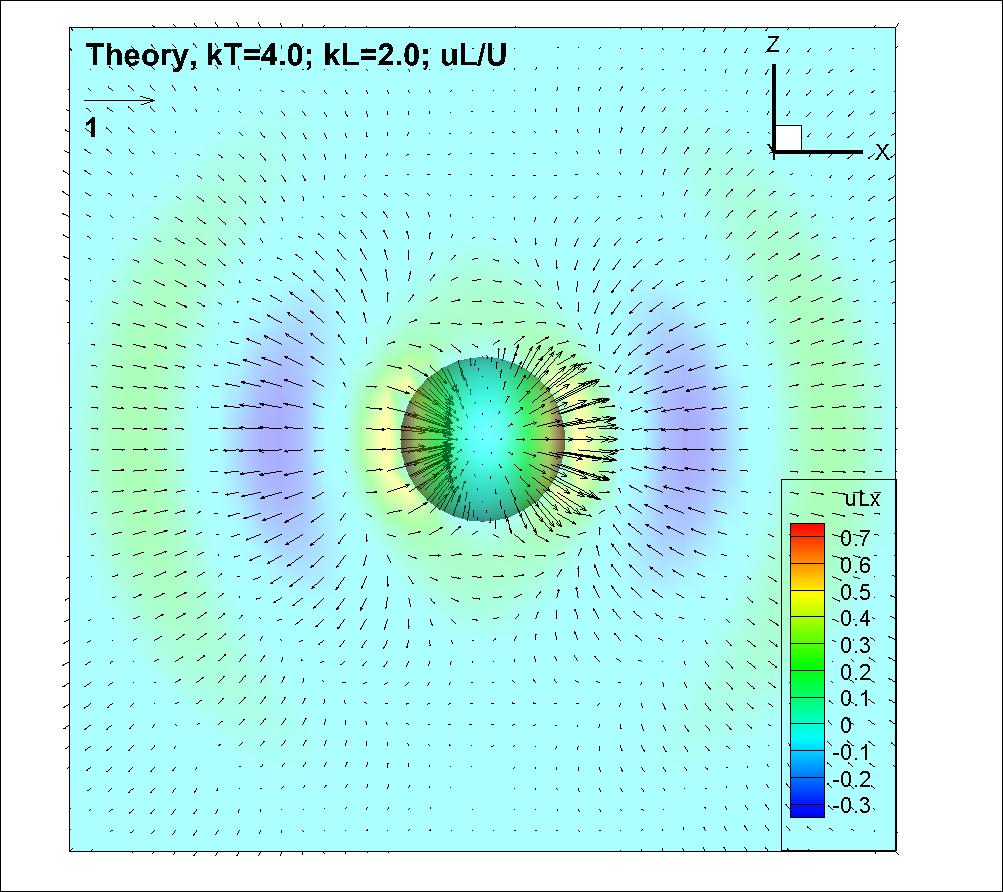}
   \includegraphics[width=0.5\textwidth]{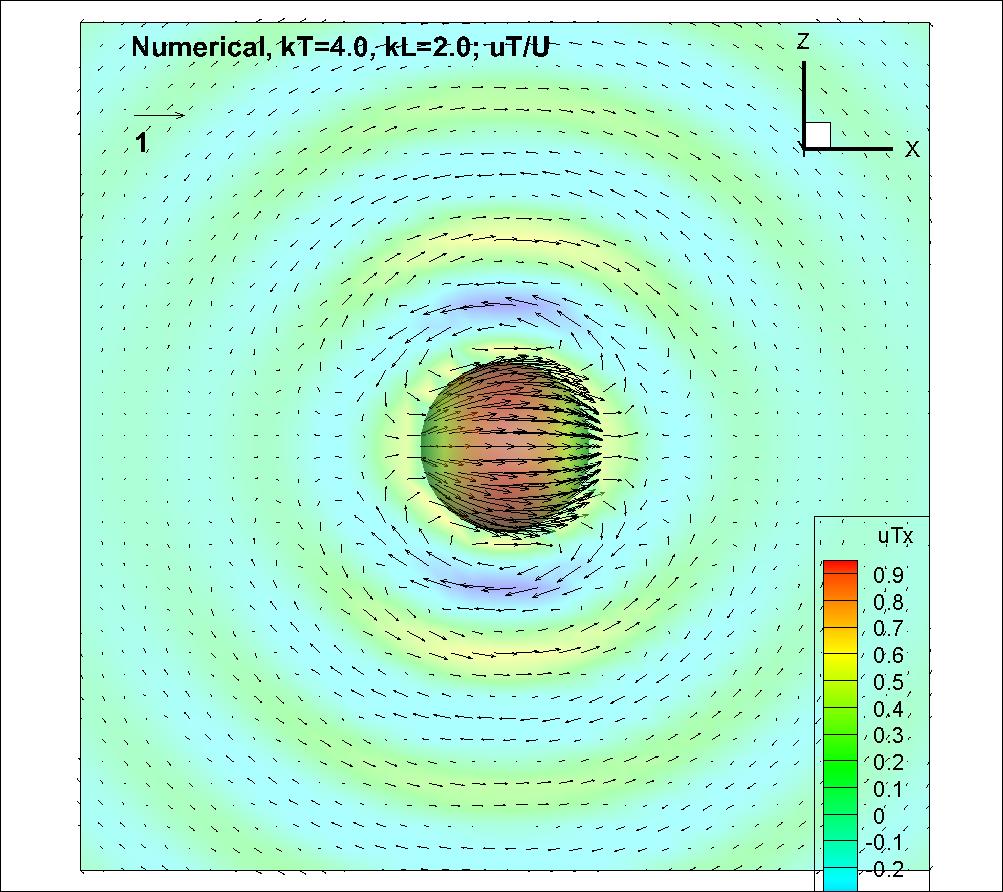}
   \includegraphics[width=0.5\textwidth]{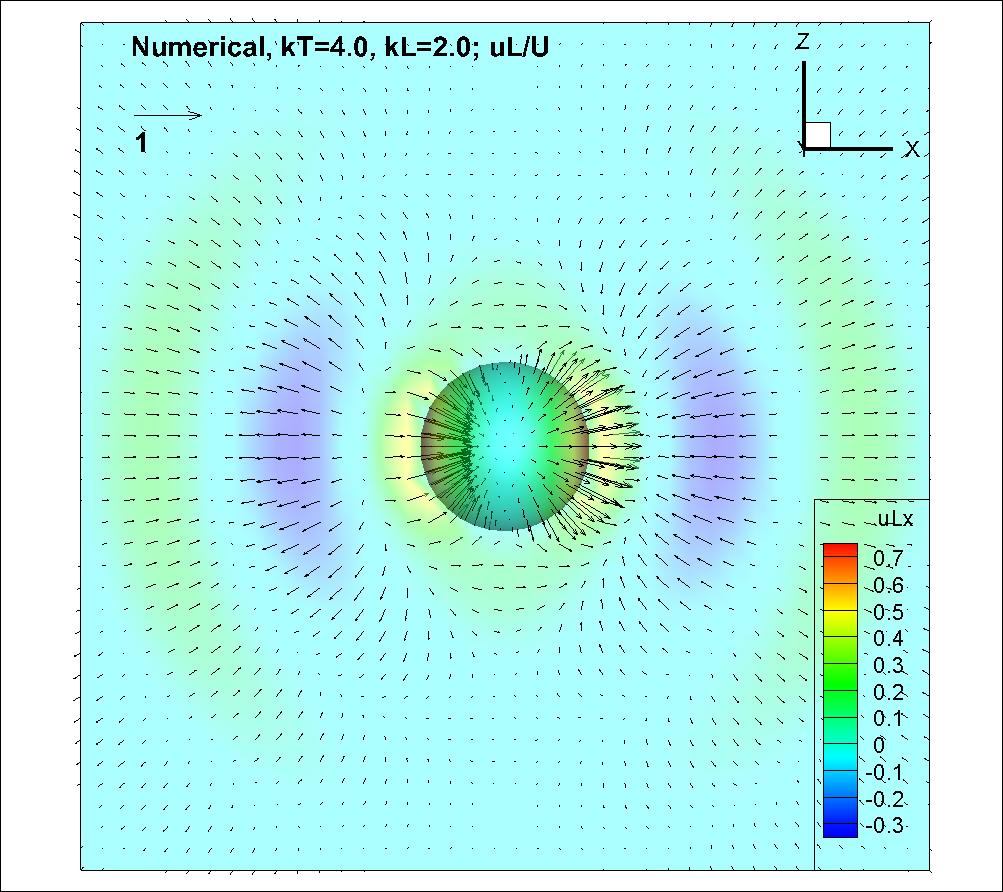}
\caption{As for Fig.\ref{fig:1} but with parameters $k_Ta=4.0$ and $k_La=2.0$. The $\boldsymbol u_T$ and $\boldsymbol u_L$ vectors are now comparable in magnitude. For corresponding movies see Sect. \ref{CompMat}.
}
\label{fig:2}       
\end{figure*}

\begin{figure*}[t]
  \includegraphics[width=0.5\textwidth]{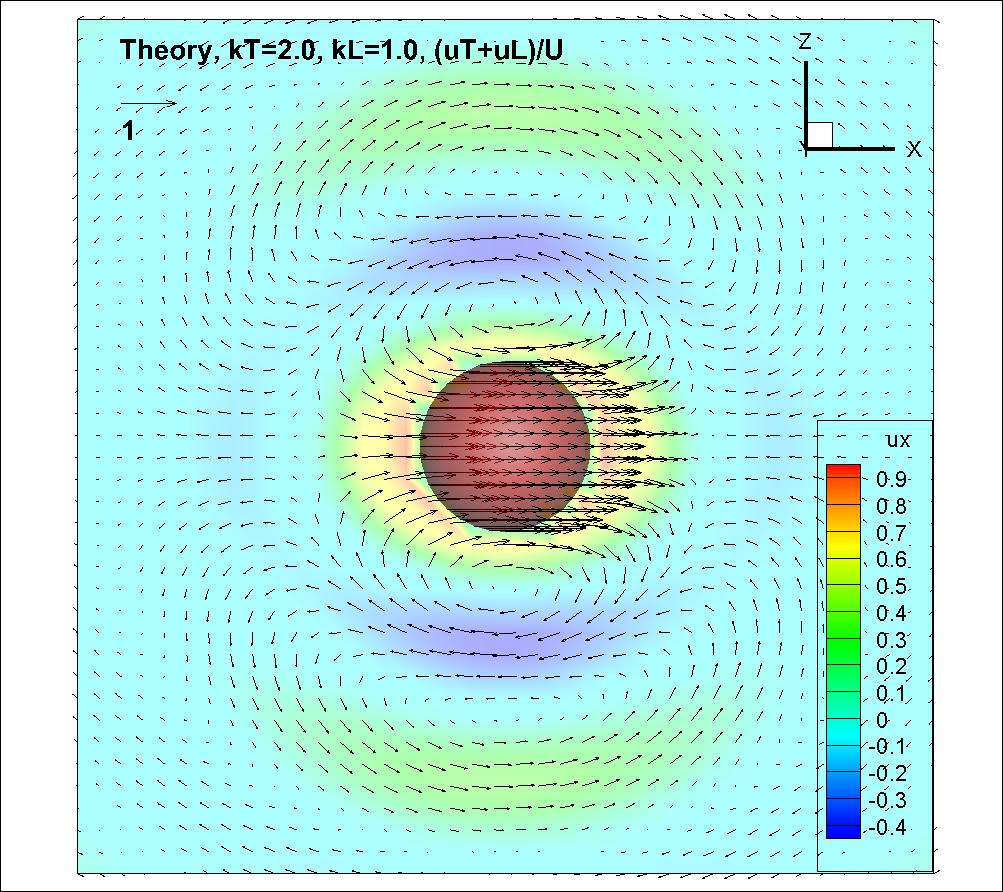}  \includegraphics[width=0.5\textwidth]{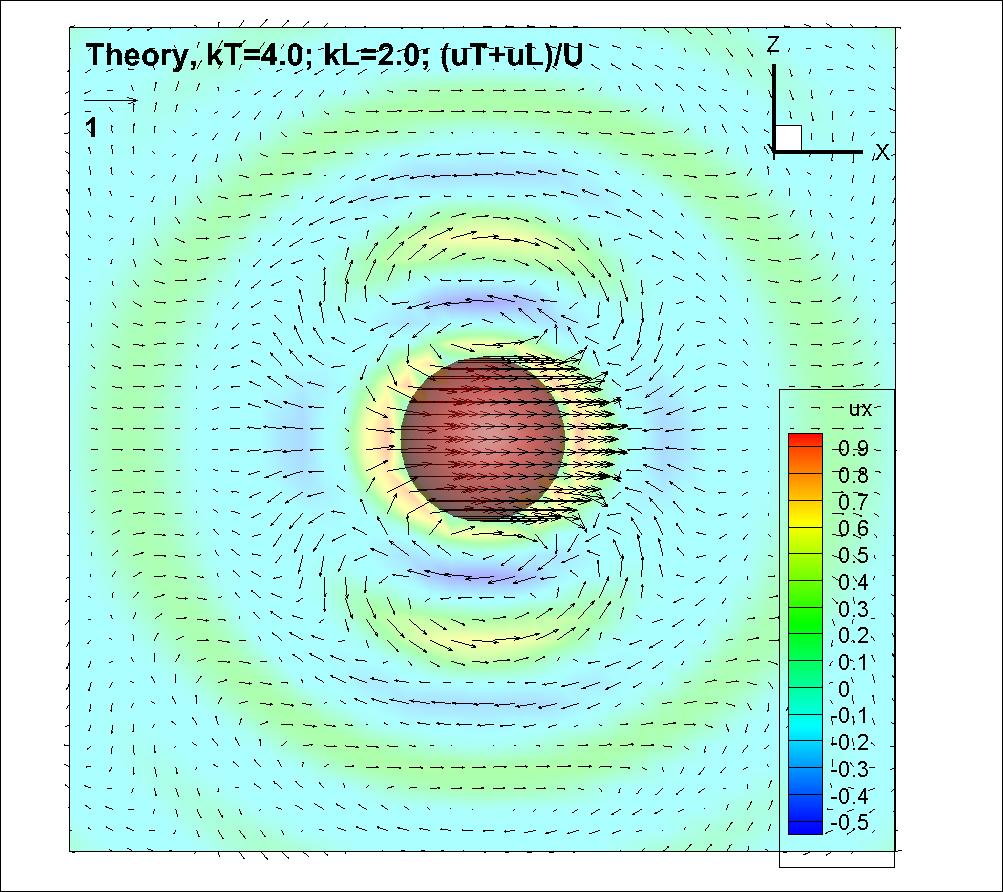}
\caption{Sphere vibrating in the $x$-direction, surface and field plot of the total theoretical $\boldsymbol u =\boldsymbol u_T + \boldsymbol u_L$ vector field; left $k_Ta=2.0$ and $k_La=1.0$; right $k_Ta=4.0$ and $k_La=2.0$. The numerical fields are virtually indistinguishable from those above (not shown). For corresponding movies see Sect. \ref{CompMat}.
}
\label{fig:3}       
\end{figure*}

\begin{figure*}[t]
  \includegraphics[width=0.2\textwidth]{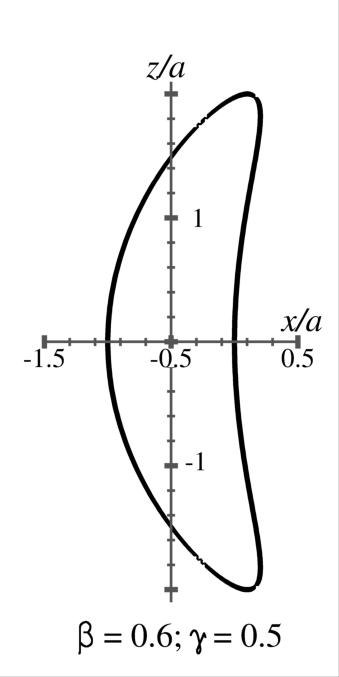} 
  \includegraphics[width=0.4\textwidth]{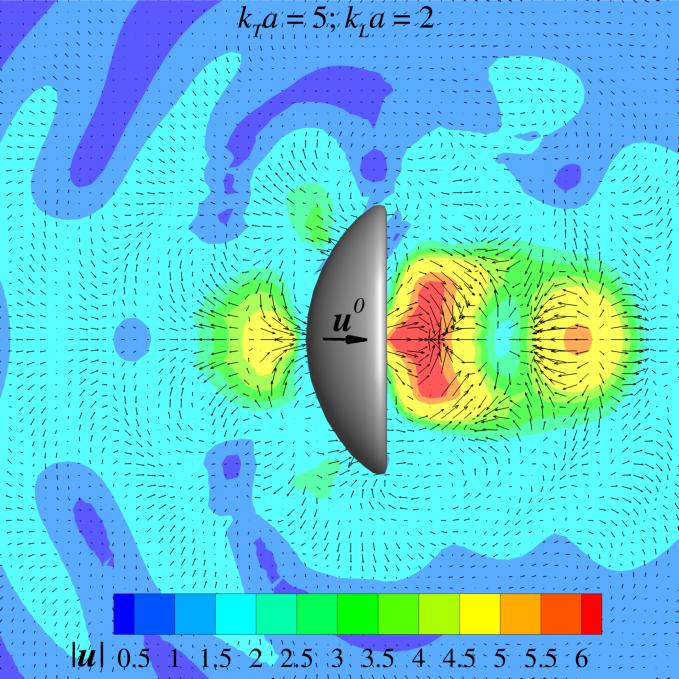}  
  \includegraphics[width=0.4\textwidth]{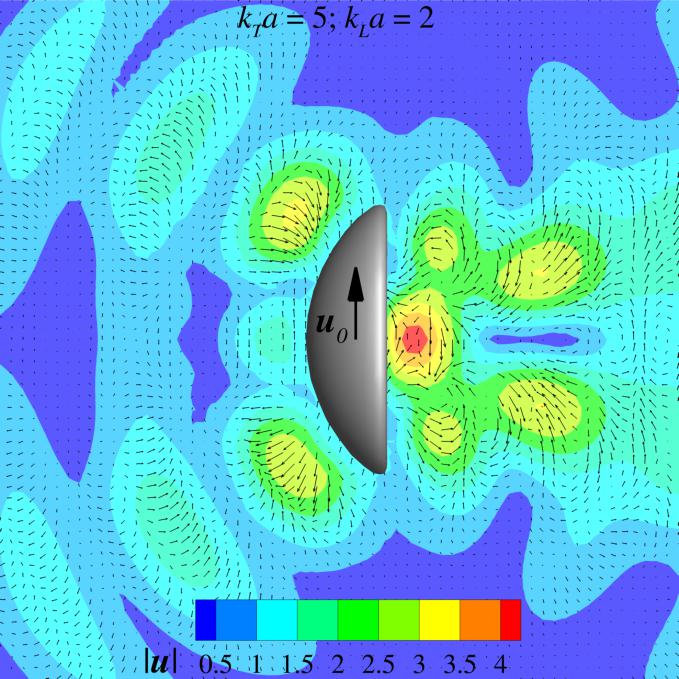}
\caption{Bowl-shaped oscillator (left) vibrating in (middle) the $x$-direction parallel to its axis of symmetry and (right) $z$-direction perpendicular to its axis of symmetry, field plot of $\boldsymbol u$ vector field scaled by $U$ obtained numerically; $k_Ta=5.0$ and $k_La=2.0$. The wave focusing effect of the bowl-shaped object can clearly be observed, even in the case of the oscillation in the $z$-direction, perpendicular to the axis of symmetry of the bowl (right). For corresponding movies see Sect. \ref{CompMat}.
}
\label{fig:4}       
\end{figure*}

Results will now be shown for the vibrating sphere with $\boldsymbol u^0 = (U,0,0)$ and numerical BEM results are compared to the analytic solution of Sect. \ref{Spheresolution}. In all examples, the sphere is represented by a mesh with 180 quadratic elements and $N=362$ nodes. The field values were obtained through post-processing on a $40\times40$ grid covering an area of $10a\times 10a$ of the 3D domain outside the sphere. In Fig. \ref{fig:1}, we compare analytic and numerical results for $\boldsymbol u_L$ and $\boldsymbol u_T$ with $k_La=2.0$, $k_Ta=1.0$. For this particular parameter set, $\boldsymbol u_T$ is the dominant term. The agreement between theory and numerical results is excellent. This can be seen clearly in Fig. \ref{fig:1_1} where the average difference between the numerical solution and the analytic solution is less than 0.13\%. Another set of comparisons with $k_Ta=4.0$ and $k_La=2.0$ is shown in Fig. \ref{fig:2} for which the $\boldsymbol u_L$ component is slightly more prominent. In Fig. \ref{fig:3}, the total field $\boldsymbol u$ is shown for both parameter sets. In Fig. \ref{fig:4}, the total field $\boldsymbol u$ for a bowl-shaped oscillator with convex and concave surfaces vibrating parallel and perpendicular to its axis of symmetry is shown. The shape of this axisymmetric bowl-shaped oscillator is obtained by rotating the following curve around the $x$-axis (see Eq. (6) in \cite{KlaseboerAO2017} and also \cite{KlaseboerAcoustics} for an application in acoustic waves) 
\begin{align}\label{eq:bowlshape}  
(x/a, \text{ } z/a) = (\beta \sin^2{\alpha} + \gamma [\cos{\alpha}-1],\text{ } 2\sin{\alpha}), \qquad 0 \leq \alpha \leq 2\pi,
\end{align}
where the parameters $\beta=0.6$, $\gamma=0.5$, $k_{T}a=5$ and $k_{L}a=2$ are chosen in Fig. \ref{fig:4}. 

Once the (complex) displacements fields: $\boldsymbol u$, $\boldsymbol u_T$ or $\boldsymbol u_L$ are obtained, we can make use of the fact that when this solution is multiplied by a constant phase factor, i.e. $\boldsymbol{u} \exp(i \alpha)$, it is also a solution of the system. This was used to reconstruct the solution in the time domain and get the solution at different time intervals. The movie files thus created are available as supplementary material. For a list of movie files see Sect. \ref{CompMat}. 


In addition to the iterative solution framework discussed in Sect. \ref{BEM}, a direct solution using a bigger matrix system was also investigated. One option is to solve directly for the 5 unknowns $\phi$, $\partial \phi/ \partial n$, $\partial u_{Tx}/\partial n$, $\partial u_{Ty}/\partial n$ and $\partial u_{Tz}/\partial n$ resulting in a matrix system which is $5N \times 5N$ in size (where $N$ is the number of nodes), here we still solve five Helmholtz equations, but now do so simultaneously without iteration. Another option is not to work with the potential representation for $\boldsymbol u_L = \nabla \phi$, but work directly with the $\boldsymbol u_L$ vector and its normal derivatives, this will result in a system of $9N \times 9N$ equations. Here, we do not recommend the above mentioned approaches for the following reasons: firstly, the matrix system is very large, resulting in rather long computational times. Secondly, the condition number of the $5N$ and $9N$ systems appears to be quite large resulting in spurious solutions for the decomposed vectors (nevertheless, the field vectors of the total displacement field appear to remain very accurate).

The advantage of the current iterative method over a full tensor description like the one used by Rizzo \emph{et al.} \cite{Rizzo1985} is that our method uses $N \times N$ matrices, while they use $3N \times 3N$ matrices (since there are three components for the displacement and traction in 3D). The current approach is also conceptually simpler than that of Rizzo \emph{et al.} \cite{Rizzo1985}, since there are no singular integrals to be considered. Moreover, with their method, one cannot get the transversal and longitudinal components which might have important physical implications since they travel at different speeds $c_L$ and $c_T$ as given by Eq. \ref{eq:DLE5}. This is apparent in earthquake science with the clear distinction between arrival times of $P$ waves and $S$ waves.


\section{Discussion: the zero frequency divergence}
\label{Discussion}

One final issue worth mentioning is the appearance of a zero frequency divergence of the decomposed displacement vectors $\boldsymbol u_L$ and $\boldsymbol u_T$. Eq. \ref{eq:Analytic1} can alternatively be written as:
\begin{equation} \label{eq:Analytic1b} 
\begin{aligned}
u_i ={} &  c_1 a U_{ij} u_j^0 - c_2 a^3 \frac{\partial^2}{\partial x_i \partial x_j} \frac{e^{ik_Lr}}{r} u_j^0.
\end{aligned}
\end{equation}
The term with $c_1$ is actually proportional to the Green's function of the dynamic linear elastic problem $U_{ij}$, and the term proportional to $c_2$ is a dipole tensor. Let us investigate the analytical solution when the frequency $\omega$ goes to zero. By doing a Taylor expansion of $e^{ix}$ to the second order i.e. $e^{ix}=1+ix-x^2/2+o(x^3)$, where $x$ is either $k_Tr$ or $k_Lr$, in the limit of the zero frequency, $k_L\to 0, k_T \to 0$, and the terms in Eq. \ref{eq:Analytic1} can be approximated by
\begin{eqnarray}
  \lim_{k_L,k_T\to 0} e^{ix} [1+G(x)]= -\frac{1}{x^2} + \frac{1}{2}, 
\label{eq:LimDivergence1}\\ 
  \lim_{k_L,k_T\to 0} e^{ix} G(x)= -\frac{1}{x^2} - \frac{1}{2}, 
 \label{eq:LimDivergence2}\\
   \lim_{k_L,k_T\to 0} e^{ix} F(x)= \frac{3}{x^2} + \frac{1}{2}.
 \label{eq:LimDivergence3}
\end{eqnarray} 
The first term with $c_1$ in Eq. \ref{eq:Analytic1} can now be approximated with
\begin{equation}
\begin{aligned}\label{eq:LimDivergence6}  
\lim_{k_L,k_T\to 0} \Big\{e^{i k_Tr} [1+G(k_Tr)] &- \frac{k_L^2}{k_T^2} e^{i k_Lr} G(k_Lr)\Big\} \\ 
=-\frac{1}{k_T^2r^2} &+ \frac{1}{2} 
-\frac{k_L^2}{k_T^2}\Big[-\frac{1}{k_L^2r^2} - \frac{1}{2}\Big]=\frac{1}{2}\Big[1+\frac{k_L^2}{k_T^2}\Big].
\end{aligned}
\end{equation}
It can be seen that the transversal (with $e^{ik_Tr}$) and the longitudinal (with $e^{ik_Lr}$) terms both diverge with $1/k_T^2$, but the singularities cancel each other out when they are summed. Similarly, the second term with $c_1$ in Eq. \ref{eq:Analytic1} now becomes
\begin{equation}
\begin{aligned}\label{eq:LimDivergence7}  
\lim_{k_L,k_T\to 0} \Big\{e^{i k_Tr} F(k_Tr) &- \frac{k_L^2}{k_T^2} e^{i k_Lr} F(k_Lr)\Big\} \\ 
&=\frac{3}{k_T^2r^2} + \frac{1}{2} 
-\frac{k_L^2}{k_T^2}\Big[\frac{3}{k_L^2r^2} + \frac{1}{2}\Big]=\frac{1}{2}\Big[1-\frac{k_L^2}{k_T^2}\Big].
\end{aligned}
\end{equation}
Again, the transversal and longitudinal terms both diverge with $1/k_T^2$ but cancel each other out. The term in Eq. \ref{eq:Analytic1} proportional to $c_2$ does not diverge. The constants $c_1$ and $c_2$ can also be expressed in the zero frequency limit as:
\begin{eqnarray}
  c_1^0 = \lim_{k_L,k_T\to 0} c_1 = 
  \lim_{k_L,k_T\to 0} \frac{-B}{DA-BC} =
  \frac{3}{4+2k_L^2/k_T^2},
 \label{eq:LimDivergence_c1} \\ 
c_2^0 = \lim_{k_L,k_T\to 0} c_2 = 
  \lim_{k_L,k_T\to 0} \frac{A}{DA-BC} =
  \frac{1-k_L^2/k_T^2}{4+2k_L^2/k_T^2}.
 \label{eq:LimDivergence_c2}
\end{eqnarray} 
Thus in the limit of $k_L,k_T \to 0$, the displacement field becomes
\begin{eqnarray}
  u_i = 
  a c_1^0 \Big[\frac{u_i^0}{r} + \frac{x_i x_j u_j^0}{r^3} +\frac{k_L^2}{k_T^2}\big(\frac{u_i^0}{r} -\frac{x_i x_j u_j^0}{r^3}\big)\Big] +a^3c_2^0\Big[\frac{u_i^0}{r^3} -\frac{3 x_i x_j u_j^0}{r^5} \Big].
 \label{eq:Lim_ui} 
\end{eqnarray} 
In Eq. \ref{eq:Lim_ui}, the first two terms, $u_i^0/r + x_i x_j u_j^0/r^3$, represent a so-called Stokeslet that is a divergence free part of the solution. The terms with $k_L^2/k_T^2$ in front represent the curl free part. The last part that is proportional to $c_2^0$ is both divergence and curl free, which makes the Helmholtz decomposition non-unique in the zero frequency case. Both Eqs. \ref{eq:Theory4} and \ref{eq:Theory5} then revert back to the Laplacian. Even though $k_L$ and $k_T$ are both zero, their ratio in Eq. \ref{eq:Lim_ui} remains finite since from Eq. \ref{eq:DLE5} one can obtain 
\begin{eqnarray}
\frac{k_L^2}{k_T^2} = \frac{c_T^2}{c_L^2} = \frac{\mu}{\lambda +2 \mu}.
 \label{eq:k_Lk_Tratio} 
\end{eqnarray} 

The fact that the transversal and longitudinal part of Eqs. \ref{eq:LimDivergence6} and \ref{eq:LimDivergence7} diverge when the frequency approaches zero poses some limitations on the proposed boundary element framework where we separated the solution into a divergence and a curl free part. Note that the Rizzo \cite{Rizzo1985} solution does not diverge in this limit since it does not use the Helmholtz decomposition to split $\boldsymbol u$ into $\boldsymbol u_T$ and $\boldsymbol u_L$ but works with the total displacement $\boldsymbol u$ and the traction instead, however, strong singularities will show up in their method at zero frequency. Since the divergence occurs in the Green's function $U_{ij}$, it is highly likely that any $\boldsymbol u_T$, $\boldsymbol u_L$ decomposition for an arbitrary object will exhibit the same singular behavior. 

Note that this divergence is unrelated to the zero frequency catastrophe encountered in certain numerical implementations of electromagnetic scattering (see for example Chew \cite{Chew2009}), since it originates there from the decoupling of the electric and magnetic field at zero frequency, whereas in the current case the cause of the divergence is the Helmholtz decomposition of the displacement field.


\section{Conclusion}
\label{Conclusions}

The dynamic linear elasticity problem was tackled by working with the displacement field, $\boldsymbol u$, using a Helmholtz decomposition. The transversal, $\boldsymbol{u}_T$ and longitudinal, $\boldsymbol{u}_L$ components were all solved with desingularized Helmholtz boundary element methods, with one scalar Helmholtz equation for the scalar potential, $\phi$ of the longitudinal part and three scalar Helmholtz equations for the three Cartesian components of the transversal part plus an additional scalar Helmholtz equation to enforce the divergence free condition of $\boldsymbol{u}_T$. To minimize the need to solve large matrix equations, this systems of 5 scalar Helmholtz equations are solved by an iterative method.

It was shown that this numerical approach is viable by comparing the results to that of an analytical solution for a vibrating sphere for two different sets of parameters with $ka$ around unity. Theoretically it was shown that the framework will fail for very low $ka$ numbers, since the transversal and longitudinal part both diverge. However, the total displacement remains well-behaved and finite. Thus the current framework works best for moderately high $ka$ numbers.   

\section{Complementary material description}
\label{CompMat}
The following movies are available as complementary material and correspond to the test cases described in the text: 
\begin{enumerate}
\item \textbf{01m\_Theory\_uTotal\_kT2\_kL1.mp4} shows the total displacement field $\boldsymbol u$ for the parameters $k_Ta=2.0$ and $k_La=1.0$. At several radii away from the sphere, the main displacement occurs around the $z$-axis in the horizontal direction. The contour plots correspond to the $x$-component of the $\boldsymbol u$ vector. 
\item \textbf{02m\_Theory\_uTotal\_kT2\_kL1b.mp4}; as the previous movie, but now the contour plots are for the z-component of the $\boldsymbol u$ vector.
\item \textbf{03m\_Theory\_uT\_kT2\_kL1b.mp4}; the same parameters as for the previous movies, but now the transversal components $\boldsymbol{u}_T$ are shown. The main transversal waves move away from the sphere along the $z$-axis. The $x$-component is shown as a contour plot. 
\item \textbf{04m\_Theory\_uL\_kT2\_kL1b.mp4}; the same parameters as for the previous movies, but now the longitudinal components $\boldsymbol u_L$ are shown (with the $x$-component again as a contour plot). The main longitudinal waves are moving along the $x$-axis.
\item \textbf{05m\_Theory\_uTotal\_kT4\_kL2.mp4} shows the total displacement field $\boldsymbol u$ for the parameters $k_Ta=4.0$ and $k_La=2.0$. Due to these higher $ka$ numbers the wavelengths are shorter. The contour plots are for the $x$-component. The overall pattern at some distance away from the sphere appears to be more `radial' in nature than for the parameters  $k_Ta=2.0$ and $k_La=1.0$. 
\item \textbf{06m\_Theory\_uTotal\_kT4\_kL2b.mp4} is the same as the previous movie, but now with the contour plot for the $z$-component. 
\item \textbf{07m\_Theory\_uT\_kT4\_kL2b.mp4}; as for the previous two movies, but now the transversal decomposed vector field $\boldsymbol u_T$ is shown. It appears to `radiate' mainly in the z-direction. 
\item \textbf{08m\_Theory\_uL\_kT4\_kL2b.mp4} as for the previous three movies, now for the longitudinal decomposed vector field $\boldsymbol u_L$. This time the waves `radiate' outwards mainly in the $x$-direction. 
\item \textbf{09m\_Bowl\_u\_kT5\_kL2\_parallel.mp4} shows the total displacement field $\boldsymbol u$ for the parameters $k_Ta=5.0$ and $k_La=2.0$ when a bowl-shaped oscillator vibrates along its axis of symmetry. The contour plots correspond to the magnitude of the $\boldsymbol u$ vector.  
\item \textbf{10m\_Bowl\_u\_kT5\_kL2\_perpendicular.mp4} shows the total displacement field $\boldsymbol u$ for the parameters $k_Ta=5.0$ and $k_La=2.0$ when a bowl-shaped oscillator vibrates perpendicular to its axis of symmetry. The contour plots correspond to the magnitude of the $\boldsymbol u$ vector.
\end{enumerate}
The movie files are best appreciated when the player is put in the ``loop'' mode. The vectors on the surface of the sphere have been suppressed in the plotting routine in order to see the vectors in the field better. 

\appendix
\section{An oscillating rigid sphere in an elastic medium}

In this Appendix, we sketch the derivation of the analytic solution that describes the periodic movement of a rigid no-slip sphere of radius, $a$ in an infinite elastic medium. This solution is inspired by the well-known analytic solution of a similar sphere in a quiescent viscous liquid at low Reynolds number or Stokes flow with the following governing equations for the velocity $\boldsymbol{u}$ and pressure $p$: $\mu \nabla^2 \boldsymbol{u} = \nabla p$ and $ \nabla \cdot \boldsymbol{u}=0 $, with $\mu$ the viscosity of the liquid.
The solution for the velocity field, in tensor notation, is:
\begin{equation} \label{eq:Stokes} 
u_i^{Stokes} =  \Big[ \frac{3a}{4r} + \frac{a^3}{4 r^3}\Big]  u_i^0  + \frac{3}{4a^2}\Big[\frac{a^3}{r^3} - \frac{a^5}{r^5}\Big] x_i (x_j u_j^0)
\end{equation}
with $u_i^0$ being the velocity of the sphere, that is, $\boldsymbol{u}=\boldsymbol{u}^0$ on the sphere surface and $\boldsymbol{u}$ decays as $1/r$ towards infinity. Integration of the corresponding traction over the surface of the sphere leads to the Stokes formula for the drag force on a sphere: $\boldsymbol{F}_d = (6\pi \mu a) \boldsymbol{u}^0$. 

Eq. \ref{eq:Stokes} can be rewritten in a more convenient form for our analysis as:
\begin{subequations} \label{eq:Stokes2} 
\begin{align}
u_i^{Stokes} =& \; \left(\frac{3a}{4}\right)\Big[ \frac{\delta_{ij}}{r} + \frac{x_i x_j}{r^3}\Big]  u_j^0
       + \left(\frac{a^3}{4}\right) \Big[\frac{\delta_{ij}}{r^3} - 3\frac{x_i x_j}{r^5} \Big]  u_j^0, \\
            \equiv& \; \left(\frac{3a}{4}\right) \quad \; G_{ij}^{Stokes} \quad  u_j^0 \; - \left(\frac{a^3}{4}\right) \quad \nabla (\nabla \frac{1}{r}) \cdot \boldsymbol{u_0}.
\end{align}
\end{subequations}
The term: $G_{ij}^{Stokes} \equiv \Big[ \frac{\delta_{ij}}{r} + \frac{x_i x_j}{r^3}\Big]$ is a Stokeslet or the Green's function for Stokes flow whereas the second term: $\frac{\partial^2}{\partial x_i \partial x_j} (\frac{1}{r}) = \nabla (\nabla \frac{1}{r})$ is the dipolar Green's function of the Laplace equation: $\nabla^2 \phi = 0$.

Now we observe that the dipolar term: $\nabla (\nabla \frac{1}{r})$ is a solution of the governing equation for \emph{static} linear elasticity:
\begin{equation} \label{eq:LinElas1}
\Big[\frac{k_T^2}{k_L^2}-1 \Big]\nabla \nabla \cdot \boldsymbol{u} + \nabla^2 \boldsymbol{u}=0    
\end{equation}
so analogous to Eq. \ref{eq:Stokes2} we seek a general solution of Eq. \ref{eq:LinElas1} of the form
\begin{equation} \label{eq:LinElas2}
u_i^{LE} =c_1 \; a \; G_{ij}^{LE}  u_j^0
       - c_2 \;a^3 \; \left(\frac{\partial^2}{\partial x_i \partial x_j} \frac{1}{r} \right) u_j^0,
\end{equation}
where $c_1$ and $c_2$ are constants to be determined and $G_{ij}^{LE}$ is the Green's for the static linear elastic equation
\begin{equation} \label{eq:GreenLE}
G_{ij}^{LE} \equiv \Big[\frac{\delta_{ij}}{r} + \frac{x_i x_j}{r^3}\Big] + \frac{k_L^2}{k_T^2}\Big[\frac{\delta_{ij}}{r} - \frac{x_i x_j}{r^3} \Big].
\end{equation}

We find the constants $c_1$ and $c_2$ using the boundary condition at $r=a$: $u_i^{LE}=u_i^0$ which leads to
\begin{equation} \label{eq:LinElas3}
 u_i^{LE}=u_i^0=u_i^0 \Big[\big\{ 1+\frac{k_L^2}{k_T^2}\big\}c_1 + c_2 \Big] + \frac{x_i x_j}{a^2}u_j^0
 \Big[\Big\{1-\frac{k_L^2}{k_T^2}\big\}c_1 - 3 c_2 \Big] \quad \text{at} \quad r=a,
\end{equation}
The second term in square brackets must be zero and the first term in square brackets must then be equal to 1. Thus solving for $c_1$ and $c_2$ results in: 
\begin{equation}
c_1=\frac{3k_T^2}{4k_T^2+2k_L^2},
\end{equation}
\begin{equation}
c_2=\frac{k_T^2-k_L^2}{4k_T^2+2k_L^2}.
\end{equation}

This approach can be extended to the \emph{dynamic} linear elastic case by taking a linear combination of the Green's function for dynamic linear elasticity and a term proportional to the Helmholtz dipole $\frac{\partial^2}{\partial x_i \partial x_j} \left( \frac{exp(ikr)}{r}\right) u_j^0$. For dynamic linear elasticity, the vector $\boldsymbol{u}$ represents the velocity amplitude of a vibrating sphere that is a constant in the frequency domain. After some algebra, this approach leads eventually to Eq. \ref{eq:Analytic1}.
%

%
%



\begin{thebibliography}{}
%
\bibitem{Iturranan2008}
Iturrar\'an-Viveros U, S\'anchez-Sesma FJ, Luz\'on F, Boundary element simulation of scattering of elastic waves by 3-D cracks, J. Applied Geophysics 64, 70-82 (2008)

\bibitem{Beskos1}
Beskos DE, Boundary element methods in dynamic analysis, Appl. Mech. Rev. 40, 1-23 (1987)

\bibitem{Dual2012}
Dual J, Schwarz T, Acoustofluidics 3: continuum mechanics for ultrasonic particle manipulation, Lab Chip 12, 244-252 (2012)

\bibitem{Sternberg1960}
Sternberg E, On the integration of the equations of motion in the classical theory of elasticity, Archive for Rational Mechanics and Analysis, 6, 34-50 (1960)

\bibitem{Gurtin1973}
Gurtin ME, The linear theory of elasticity, Springer-Verlag Berlin Heidelberg (1973)

\bibitem{CruseRizzo1968}
Cruse TA, Rizzo FJ, A direct formulation and numerical solution of the general transient elastodynamic problem.I, J.Math.Analysis and Applications 22, 244-259 (1968)

\bibitem{Cruse1968}
Cruse TA, A direct formulation and numerical solution of the general transient elastodynamic problem.II, J.Math.Analysis and Applications 22, 341-355 (1968)

\bibitem{Rizzo1985}
Rizzo FJ, Shippy DJ, Rezayat M, A boundary integral equation method for radiation and scattering of elastic waves in three dimensions, Int.J.Numerical Methods in Engineering 21, 115-129 (1985)

\bibitem{Beskos2}
Beskos DE Boundary element methods in dynamic analysis: part II (1986-1996), Appl. Mech. Rev. 50, 149-197 (1997)

\bibitem{Bu2014}
Bu, F, Lin, J, Reitich F, A fast and high-order method for the three-dimensional elastic wave scattering problem, J.Comp.Physics 258, 856-870 (2014)

\bibitem{SunRoySoc2015}
Sun Q, Klaseboer E, Khoo BC, Chan DYC, Boundary regularized integral equation formulation of the Helmholtz equation in acoustics, R. Soc. Open Sci. 2, 140520 (2015)

\bibitem{KlaseboerEABEM2009}
Klaseboer E, Rosalez-Fernandez C, Khoo BC, A note on true desingularization of boundary element methods for three-dimensional potential problems, Engng. Anal. Bound. Elem. 33, 796-801 (2009) 

\bibitem{LandauLifshitz}
Landau LD, Lifshitz EM, Theory of Elasticity, Pergamon Press Ltd, Oxford (1959)

\bibitem{Harrington2001}
Harrington RF, Time-harmonic electromagnetic fields, page 38, John Wiley \& Sons, Inc. New York (2001)

\bibitem{KlaseboerIEEE2017}
Klaseboer E, Sun Q, Chan DYC, Nonsingular field-only surface integral equations for electromagnetic scattering, IEEE Transactions on Antennas and Propagation 65 972-977 (2017)

\bibitem{SunPRB2017}
Sun Q, Klaseboer E, Chan DYC, Robust multiscale field-only formulation of electromagnetic scattering, Physical Review B 95 045137 (2017)

\bibitem{Lautrup}
Lautrup B, Physics of continuous matter, 2nd Edition (2011) CRC Press, Taylor and Francis,  Sect. 12.6, page 204 (2004)

\bibitem{KlaseboerJFM2012}
Klaseboer E, Sun Q, Chan DYC, Non-singular boundary integral methods for fluid mechanics applications, J.Fluid Mechanics 696, 468-478 (2012)

\bibitem{Becker1992}
Becker AA, The boundary element method in engineering: a complete course, McGraw-Hill International (UK) Limited (1992)

\bibitem{Kirkup1998}
Kirkup S, The boundary element method in acoustics ISBN 0 953 4031 06 (1998)

\bibitem{KlaseboerAO2017}
Klaseboer E, Sun Q, Chan DYC, Field-only integral equation method for time domain scattering of electromagnetic pulses, Appl. Optics 56, 9377-9383 (2017).

\bibitem{KlaseboerAcoustics}
Klaseboer E, Sepehrirahnama S, Chan DYC, Klaseboer E,  Space-time domain solutions of the wave equation by a non-singular boundary integral method and Fourier transform, J. Acoust. Soc. Am. 142, 697-707 (2017).

\bibitem{Chew2009}
Chew WC, Tong MS, Hu B, Integral equation methods for electromagnetic and elastic waves, Morgan \& Claypool (2009).




\end{thebibliography}


\end{document}